\documentstyle[11pt,newpasp,twoside,psfig]{article}
\index{summary}
\index{instructions}
\index{template}

\def\edcomment#1{\iffalse\marginpar{\raggedright\sl#1\/}\else\relax\fi}
\reversemarginpar

\begin{document}

\title{SNR G292.2$-$0.5: A Radio Supernova Remnant Associated with the Young
Pulsar J1119$-$6127}

\author{F. Crawford}
\affil{Lockheed Martin Management and Data Systems, P.O. Box 8048,
Philadelphia, PA 19101, USA}

\author{B. M. Gaensler}
\affil{Harvard-Smithsonian Center for Astrophysics, 60 Garden Street,
Cambridge, MA 02138, USA} 

\author{V. M. Kaspi}
\affil{Department of Physics, McGill University, 3600 University Street,
Montreal, Quebec, H3A 2T8, Canada}

\author{R. N. Manchester}
\affil{ATNF, CSIRO, P.O. Box 76, Epping, NSW 1710, Australia}

\author{F. Camilo} 
\affil{Columbia Astrophysics Laboratory, Columbia University,
550 West 120th Street, New York, NY 10027, USA} 

\author{A. G. Lyne} 
\affil{University of Manchester, Jodrell Bank Observatory, Macclesfield,
Cheshire SK11 9DL, UK} 

\author{M. J. Pivovaroff}
\affil{Space Sciences Laboratory, University of California, Berkeley,
260 SSL \#7450, Berkeley, CA 94720-7450, USA}

\begin{abstract}
We report on Australia Telescope Compact Array observations in the
direction of the young high-magnetic-field radio pulsar PSR
J1119$-$6127. In the resulting images we identify a non-thermal radio
shell of diameter $15\arcmin$, which we classify as a previously
uncataloged young supernova remnant (SNR), G292.2$-$0.5. This SNR is
positionally coincident with PSR J1119$-$6127, and we conclude that
the two objects are physically associated. No radio emission is
detected from any pulsar wind nebula (PWN) associated with the pulsar;
our observed upper limits are consistent with the expectation that
high-magnetic-field pulsars produce radio nebulae which fade
rapidly. This system suggests a possible explanation for the lack of
associated radio pulsars and/or PWNe in many SNRs.
\end{abstract}

\section{Introduction: PSR J1119$-$6127}

PSR J1119$-$6127 is a 408-ms pulsar with a large inferred surface
magnetic field strength ($B \sim 4 \times 10^{13}$ G) which was
discovered in the Parkes Multibeam Pulsar Survey (Camilo et
al. 2000). A measured $\dot{P}$ and $\ddot{P}$ for the pulsar
indicates an age of $\tau = 1.7 \pm 0.1$ kyr (assuming an initial spin
period $P_{0} \ll P$), making this the youngest pulsar discovered in
the survey. Despite its similarity in age to the Crab pulsar, PSR
J1119$-$6127 has significantly different characteristics, most notably
its much larger period and magnetic field.  Since it is young, PSR
J1119$-$6127 is a good candidate to have an associated supernova
remnant (SNR) and pulsar wind nebula (PWN).  These can be detected and
studied using radio imaging.

\section{Radio Imaging of PSR J1119$-$6127 with ATCA} 

We have observed the region containing PSR J1119$-$6127 with the
Australia Telescope Compact Array (ATCA) radio interferometer.  Both
1.4 and 2.5 GHz data were taken using pulsar gating, and data were
combined from multiple array configurations.  A 1.4 GHz total
intensity map is shown in Figure 1.  A limb-brightened radio shell of
$\sim 15\arcmin$ diameter is evident, which we call
G292.2$-$0.5. Using a spectral tomography technique to compare the
flux in the 1.4 and 2.5 GHz maps (Katz-Stone \& Rudnick 1997; Crawford
2000), we measured a spectral index for the brightest part of the
shell of $\alpha = -0.6 \pm 0.2$ ($S \sim \nu^{\alpha}$).  Off-pulse
maps from pulsar gating indicate no detectable PWN emission coincident
with the pulsar.  We quantified these upper limits by adding synthetic
PWNe of varying sizes and fluxes to the data prior to mapping to
determine the threshold of detectability. No polarization information
from the observations could be used owing to uncorrectable
instrumental leakage from bright nearby H{\sc ii} regions.

\section{Results}

\subsection{G292.2$-$0.5: A New Shell SNR} 

The measured $\alpha = -0.6 \pm 0.2$ for the shell indicates a
non-thermal (synchrotron) origin for the radio emission, which is
expected for a SNR. There is no significant infrared emission
coincident with the shell, and the observed spectral index is
consistent with those measured for other young shell SNRs (Table
1). We conclude from this evidence that G292.2$-$0.5 is a previously
uncataloged radio shell SNR.

\subsection{An Association Between PSR J1119$-$6127 and 
SNR G292.2$-$0.5}

The circular, limb-brightened morphology of the SNR (Figure 1) is
expected for a young SNR which has not yet been deformed by the
interstellar medium (e.g., Cas A, Kepler) and suggests the SNR is
young. The location of PSR J1119$-$6127 at the center of shell is also
consistent with an association with the pulsar: the probability of a
chance alignment within 1$'$ of the shell center is estimated to be
$\la 10^{-4}$. This positional coincidence is also consistent with the
youth of the pulsar. If the pulsar were born at the shell center, then
the implied transverse velocity is reasonable ($v_{\rm PSR} \la 500$
km s$^{-1}$) for a 5 kpc SNR distance. The implied shell size in this
case is $R_{\rm shell} \sim 10$ pc, and for an assumed SNR age of
$\tau = 1.7$ kyr, this corresponds to a free expansion velocity of
$v_{\rm shell} \sim 6000$ km s$^{-1}$. These $R_{\rm shell}$ and
$v_{\rm shell}$ values are consistent with other SNR shells of similar
age (Table 1). From this evidence, we conclude that PSR J1119$-$6127
and SNR G292.2$-$0.5 are physically associated.

\subsection{Absence of an Observable PWN}

We can use the observed upper limits on PWN emission to test an
evolutionary model of PWNe proposed by Bhattacharya (1990).
Qualitatively, the model suggests that a high-magnetic-field pulsar
born spinning rapidly undergoes severe magnetic braking at early
times.  The bulk of the rotational energy of the pulsar is quickly
deposited into the nebula, and the pulsar cannot continue to
significantly power the PWN at later times.  The PWN rapidly fades
from expansion losses and is not detectable at $\tau \sim 2$ kyr,
while the pulsar has slowed to a much longer period.  We have derived
a scaling formula for the predicted surface brightness $\Sigma$ of a
PWN powered by PSR J1119$-$6127 which includes many assumptions
(Reynolds \& Chevalier 1984; Crawford et al.\ 2001 and references
therein). Using this formula, we have scaled the observed Crab nebula
surface brightness to obtain a predicted value for PSR J1119$-$6127 of
$\Sigma \sim 6 \times 10^{-22}$ W m$^{-2}$ Hz$^{-1}$ sr$^{-1}$ at 1
GHz, which is well below our sensitivity limits.  There is a strong
dependence of $\Sigma$ on PWN age in the model, and our upper limits
indicate that if $P_{0} \ga 200$ ms for PSR J1119$-$6127
(corresponding to $\tau \la 1.2$ kyr), the PWN would be detectable in
our observations. The absence of a detectable PWN argues in favor of
$P_{0} \la 200$ ms for PSR J1119$-$6127 and supports the evolutionary
model of Bhattacharya (1990).

\begin{table} 
\caption{Young Shell SNR Parameters.}
\begin{tabular}{llccrl} 
\tableline
SNR & Other & $\tau$ & $R_{\rm shell}$ & $v_{\rm shell}$~~~ & $\alpha_{\rm shell}$ \\
    & Name  & (kyr)  & (pc)            & (km s$^{-1}$)          &          \\
\tableline
1987A           &                   & 0.01                 & $\sim$ 0.2 & $\sim$ 18000 & $-$0.9   \\
G111.7$-$2.1    & Cas A             & 0.3                  & 2.04       & 6700         & $-$0.77   \\ 
G4.5$+$6.8      & Kepler            & 0.4                  & 1.92       & 4700         & $-$0.64   \\ 
0540$-$69.3     &                   & 0.8                  & 7.27       & 9100         & $-$0.41   \\ 
G29.7$-$0.3     & Kes 75            & $\sim$ 1             & $\sim$ 8   & $\sim$ 8000  & $-$0.7    \\
G320.4$-$1.2    & MSH 15$-$5{\it 2} & 1.7                  & $\sim$ 20  & $\sim$ 11500 & $-$0.5    \\ 
{\bf G292.2$-$0.5}  &               & {\bf 1.7}            & {\bf $\sim$ 10}   & {\bf $\sim$ 5900}   & {\bf $-$0.6}  \\  
G332.4$-$0.4    & RCW 103           & $\sim$ 2             & $\sim$ 5   & $\sim$ 2500  & $-$0.5    \\
G260.4$-$3.4    & Puppis A          & 3.7                  & 35.2       & 9300         & $-$0.5    \\
\tableline
\tableline
\end{tabular} 
\end{table} 

\begin{figure}[t] 
\centerline{\psfig{figure=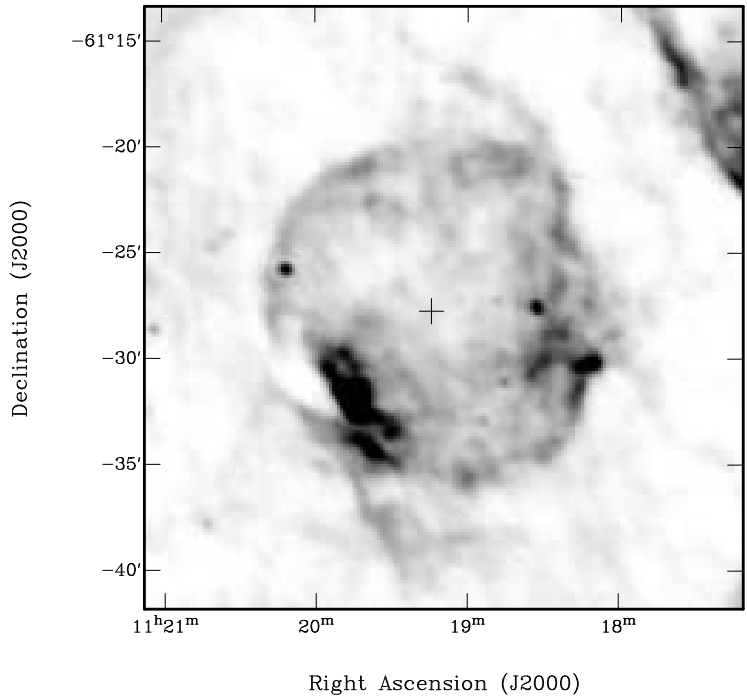,width=4.0in}} 
\caption{1.4 GHz total intensity ATCA image of the shell SNR
G292.2$-$0.5 using all baselines shorter than 7.5 k$\lambda$. The
position of PSR J1119$-$6127 is indicated by the cross.  No radio PWN
is evident.}
\end{figure} 

\section{Conclusions} 

Using ATCA radio imaging observations, we have discovered
G292.2$-$0.5, a previously unknown shell SNR with a diameter of
$15\arcmin$ and a circular, limb-brightened morphology. Evidence
strongly suggests that the SNR is associated with the young
high-magnetic-field pulsar PSR J1119$-$6127. We have detected no PWN
powered by PSR J1119$-$6127 and have quantified the upper limits on
PWN radio emission.  Our upper limits support a model of PWNe powered
by high-magnetic-field pulsars and indicate $P_{0}
\la 200$ ms for PSR J1119$-$6127. These results may provide an
explanation for the absence of detected radio pulsars and PWNe in many
shell SNRs: shells might contain high-magnetic-field pulsars which are
faint or beaming away and do not power observable PWNe. However, this
explanation depends upon the feasibility of a large population of
young high-magnetic-field pulsars. A more complete and detailed
treatment of the work described here is presented in Crawford et
al. (2001).

\end{document}